# The future of VLBI has begun!

**Huib Jan van Langevelde**[1]

*The Joint Institute for VLBI in Europe*
*Postbus 2, 7990 AA Dwingeloo, the Netherlands*
*And Sterrewacht Leiden, Leiden University*
*Postbus 9513, 2300 RA Leiden, the Netherlands*
*E-mail:* `langevelde@jive.nl`

With the exceptional progress e-VLBI has achieved over the last three years, the VLBI of the future has already started. At least for the EVN, it is argued that at some point all VLBI operations should be done in e-VLBI mode. This ambition is based on the scientific case that is described in the EVN2015 science vision. At the same time, it should be taken into account that the long-term future of radio astronomy is connected to the development of the SKA. The consensus in the community is that there is a scientific case for Very Long Baseline Interferometry in the next decade, and synergy with the technology development for the SKA and its pathfinders should be explored to enhance the VLBI capabilities. It is noteworthy that e-VLBI has been recognised as a SKA pathfinder. Here, I review the progress with e-VLBI, and the options to enhance the sensitivity and operational efficiency of the EVN and global VLBI arrays, including the options for future correlators. In the coming years, through the new NEXPReS effort, new ways are about to get introduced to enhance e-VLBI operations further to the level that all experiments can benefit from an e-VLBI component.

# 1.Introduction

VLBI is undergoing a silent revolution by providing real-time results, thereby enabling new science, boosting the reliability and potentially enhancing the sensitivity. This provides a natural growth path for high-resolution radio astronomy. And at the same time radio astronomers around the world are developing the Square Kilometre Array (SKA) [2][9], which will presumably take radio astronomy into a whole different sensitivity domain. Certainly during the construction, but probably also during operations of the first two (lower frequency) SKA phases, a continuing scientific need will exists for high resolution VLBI capabilities. The fantastic progress in the European VLBI Network, catalysed by the EXPReS[1] project, has demonstrated that e-VLBI can be the way to provide this scientific instrumentation in the next decades [11]. Below I will discuss this in detail and focus on the next steps on this roadmap to the VLBI of the future, a development that will be synergetic with the technical development of the SKA. Inevitably, I have often taken a European VLBI Network (EVN) centred perspective in this discussion.

## 2. Scientific motivation and the role of the SKA

The EVN is developing a long-term roadmap for the next decade. As funding agencies are defining scientific priorities on a European scale, the VLBI community must set its priorities in view of the global ambition of radio astronomers all over the world to develop the SKA. A while ago, the EVN has developed and published its science vision: EVN2015 (http://www.evlbi.org/publications/publications.htm).

The EVN science case is addressing a breadth of astronomical topics, ranging from the classical area of jets in Active Galactic Nuclei (AGN) and the AGN/starburst connection in cosmological fields, which calls for a boost in sensitivity. More capabilities at higher frequency are also in demand, for example to probe jet physics close to the black hole event horizon. More coverage at 22GHz and 43GHz will also stimulate the research in masers, where VLBI uniquely offers a view on the dynamics on the small scales that are relevant for star and planet formation [14]. From the same observations usually also very fundamental distance measurements are obtained.

In Galactic astronomy, the study of transient radio sources has already taken advantage of the development of real-time VLBI [12], but currently the samples under study are limited in size, until much better sensitivity becomes available. Together with large-scale survey instruments, like LOFAR [13], providing many more triggers, this can be expected to be a booming industry in a matter of years. Pulsar capabilities must feature prominently as well in the coming years, as VLBI pulsar astrometry will be in demand for population studies that harvest the results from LOFAR [4], and a number of SKA pathfinders [7][1][15]. Finally, it has been demonstrated that VLBI offers very interesting scientific applications for planetary Space missions, by providing accurate positions of spacecraft. Here too, real time processing is an important asset for the VLBI system of the future.

---

[1] EXPReS is an Integrated Infrastructure Initiative (I3), funded under the European Commission's Sixth Framework Programme (FP6), contract number 026642.

In all the science areas the VLBI sensitivity must be improved to maintain its complementary role with the EVLA [5] and e-MERLIN [3] facilities by providing high-resolution diagnostics at a similar sensitivity. Besides increases in bandwidth, sensitivity improvements are also expected from expanding the number of antennas that participate in VLBI. Indeed various new antennas are planned in Europe and beyond. A next generation correlator must provide capacity for processing these higher bandwidths, but also more flexibility for spectral line science, astrometry, pulsar binning and many bit processing.

In the EVN 2015 document, a strong science case is presented that addresses fundamental questions in astronomy. This calls for long baseline capabilities that are currently not completely covered in the SKA roadmap, certainly not in its first two phases. The VLBI science case therefore focuses on the use of Global baselines and frequencies of 5 GHz and higher. Building on the existing international collaborations in Global VLBI and EXPReS, this array could have much of its coverage in the Northern hemisphere and encompass the current EVN telescopes. A more flexible global VLBI array could be envisioned that can be deployed in tailor-made configurations for the best science.

Obviously, it will be a challenge to let this Global VLBI array take shape during the definition and construction of the SKA. The funding requirements for the SKA will demand that the radio-astronomy community set its priorities. However, in technology development there are various options to benefit directly from the SKA R+D. The VLBI facilities also play an important role for training and education and enable cutting edge astronomy research in the years to come. Moreover, the national VLBI telescopes have a unique role in keeping the public aware of the fascinating science that can be done with radio astronomy.

## 3. Progress on e-VLBI

### 3.1 EXPReS programme

In the last few years e-VLBI has evolved from an experimental technique, connecting a small number of telescopes at modest bandwidth into an operational astronomical service with competitive sensitivity and imaging capabilities. The project has been supported by the EC through the FP6 Integrating activity EXPReS, which has stimulated a large-scale collaboration between EVN technical staff and European Research Network providers.

The initial argument for developing e-VLBI has been the desire to use VLBI on transient phenomena, being able to access the data on variable sources on the timescale that they vary. Indeed, opening the parameter space accessible to VLBI has turned up some very interesting results and various new science themes have been presented [8].

However, the project has done more than enabling new science. It has also clearly demonstrated that e-VLBI is not a technique limited to do rapid response science [10]. Because of its real-time nature it is also much more robust against failure, which can be noticed immediately and addressed instantaneously. With dedicated connections this has proven to be true, even though the high-speed links potentially add an extra layer of complexity. The various demonstrations have also taken away the scepticism whether e-VLBI could ever work for intercontinental baselines, showing fringes between telescopes on different continents simultaneously and routinely.

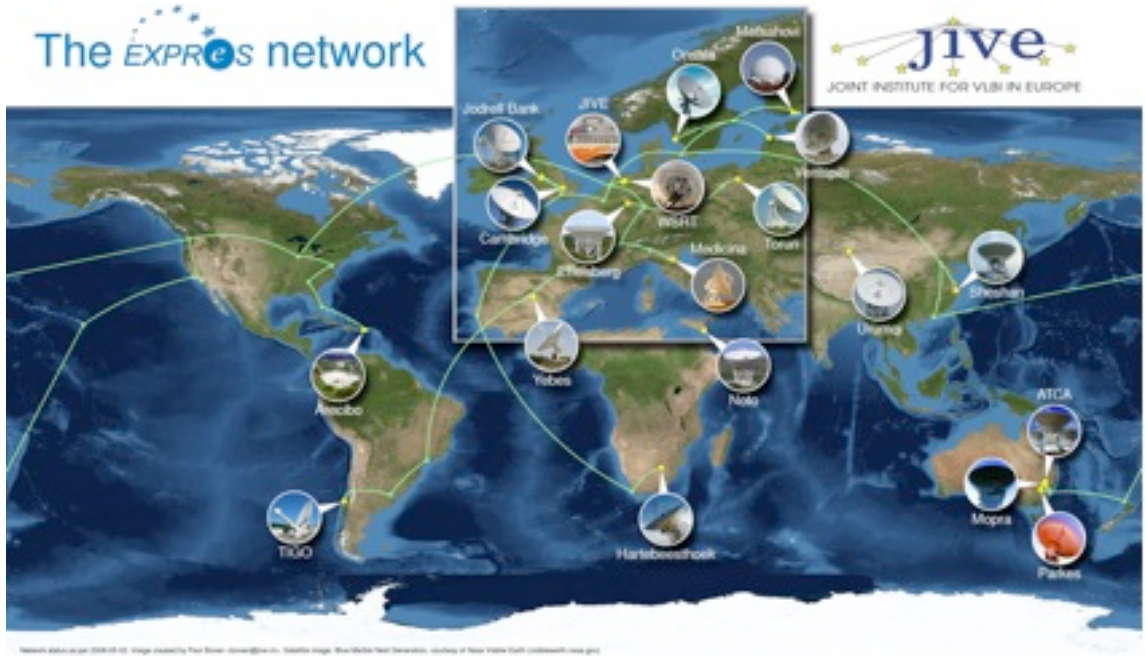

*Telescope and correlator configuration in the EXPReS network. Real-time connectivity to these telescopes was established during the course of the project.*

During the project, the VLBI community has demonstrated the power of dedicated connectivity for this application, and this kind of connectivity should be more dynamically addressable in the future. Next, the connectivity promises to be able to deliver larger and larger bandwidth with the 10Gbps standard already operational in many places. Besides reducing the delivery time to the astronomer, e-VLBI can also help in enhancing the flexibility of the observations by avoiding the complex logistics that are involved in shipping scarce recording media. Currently e-VLBI is realised in a large-scale collaboration with the connectivity providers and therefore it is hard to evaluate the economics of connected VLBI in comparison to the transport of magnetic media. Clearly in the long run it is expected that real-time connectivity will be cheaper and more environmentally friendly.

One way in which e-VLBI is less flexible than recorded VLBI is that its current operations do not allow re-correlations, which are for example needed to accommodate large numbers of telescopes or spectral line experiments that need mixed bandwidths. It is noted that in principle all of these are limitations of the current correlator, but even so it will be useful to introduce buffering of the VLBI data in various stages in order to combine the best aspects of recorded and real-time operations.

**3.2 NEXPReS programme**

The next step envisioned in the development of the VLBI of the future is to implement an e-VLBI component for every VLBI experiment that is carried out. By implementing transparent buffering mechanisms at telescope and correlator end, one can address all the current and future bottlenecks in e-VLBI. We envision an operational model in which all data that can be transported in real-time are correlated on the fly. Ideally such an e-VLBI experiment is completed directly after the observation, but there can be various logistic, technical and even scientific reasons, to access the data that was buffered and process the experiment a second time, possibly based on the evaluation of the real-time output. This way one could overcome limited connectivity to an essential station, or failures in streaming or correlation. Also reprocessing with different correlation parameters or new astronomical information on, for example source position, would be possible. Before any reprocessing is scheduled, one would have the possibility to assess the scientific need, based on the preliminary, real-time results. Such a scheme would also end almost all transportation of physical recordings, as buffering capacity on both ends could accommodate all scenarios. Such a scheme would also be more efficient in total disk capacity than the current model in which the recoding pool is fragmented to accommodate complex logistics. Implementation requires high-speed parallel recording hardware (many GB/s), as well as software systems that hide the bookkeeping from the VLBI components and their operators. A project to develop the above operations scheme has been favourably reviewed by the EC and implementation is progressing under the name NEXPReS (Novel Explorations Pushing Robust e-VLBI Services)[2].

**4. Technology development**

Recently the full 1024Mbps data-rate has been reached for e-VLBI observations [11]. The EVN has the ambition to push the sensitivity to 4 Gbps in 2010. Besides the e-VLBI effort described above and the correlator aspects discussed below, the most critical item seems to be the introduction of the new filters and digitizers, the DBBC system in the case of the EVN. This system is needed for any new telescope that joins the array and eventually for all stations to go beyond 1024Mbps. Moreover, the EVN bandwidth is limited by the inherent width of the IF systems of the telescopes, and this must be addressed at many stations in the future.

Eventually, there is the ambition to go to even higher bandwidths, which will be mostly useful at frequencies above 10 GHz. Here possibly up to 2 GHz bandwidth could be realized in each polarization, calling for at least 16 Gbps data-rates. At GHz frequencies very high data-rates can be useful for having up to 8-bit signal representation, which can overcome some of the adverse effects introduced by radio-frequency interference, notably at L-band.

The future also dictates that the existing VLBI arrays become much more flexible in scheduling and operations. Operational models must evolve in such a way that users can easily define the optimal array in extent, frequency and sensitivity for their observations, even on short notice. It should be possible to flexibly schedule and adapt observations in order to reach a guaranteed level of quality, taking for example atmospheric conditions into account.

---

[2] NEXPReS is an Integrated Infrastructure Initiative (I3), funded under the European Union Seventh Framework Programme (FP7/2007-2013) under grant agreement n° RI-261525.

**4.1 Next Generation correlator**

The current EVN Mk4 correlator at JIVE that is at the heart of the EVN operations is capable of processing 16 stations at 1024 Mbps. Even though its capacities are still being enhanced, it is already necessary to invoke multiple pass correlation for a considerable fraction of experiments. This is required for large global experiments that use more than 16 stations at one time, or for high spectral resolution; notably this occurs when continuum sensitivity is required at the same time, for example in phase referencing. Similarly, higher than 1024 Mbps data rates can in principle be supported by multiple pass correlation, however, it is obviously not possible to accommodate high data-rates for e-VLBI in this way, as long as the data is not buffered at the correlator.

Various VLBI arrays around the world are adopting software correlators that can perform VLBI processing on commodity computer clusters [6]. This approach is superior in flexibility and accuracy over hardware implementations and on moderately sized clusters these implementations perform similarly as the current hardware-based machines. At JIVE an independent software correlators has been developed in the context of the SCARIe/FABRIC programme and it is now operational at JIVE for specific VLBI projects.

For the longer-term future a large new EVN data processor is required. Looking at the above specifications and the EVN (and global) ambitions to have more stations, one can anticipate the need to process 32 stations (a factor 4 in baselines compared to the current), 16 Gbps per second (a factor 16) and maybe a 4 to 8 times better spectral resolution than the current EVN data processor. Overall the aim for 2015 must be to develop a correlator that is at least a hundred times more powerful than the current data processor. This calls for a machine of that needs to perform in a similar regime as the correlators for SKA precursors (MeerKAT, ASKAP), and some of the other pathfinders (e-MERLIN, EVLA, APERTIF). Taking future energy constraints into consideration, JIVE is actively pursuing for correlation on FPGA based architectures; a similar conclusion as was reached by other SKA pathfinder projects.

**4.2 Telescopes**

One of the main motivations to increase the correlator capacity is the increasing number of telescopes. In global observations the number of telescopes can already go beyond 16. In the near future it will be possible to observe with EVN+MERLIN, using all stations. Besides the recently added Yebes 40m (Spain) and Ventspils 32m (Latvia) dishes, the Sardinia Radio Telescope (64m) recently had its dish lifted on the bearings. And in China the new telescopes of Miyun (50m) and Kunming (40m) have participated in the Chang'e experiments. These new, big dishes will contribute to the sensitivity; in particularly for spectral line the importance of pure collecting area is enormous. In addition to newly constructed telescopes, the Russian telescopes in the KVASAR network have recently joined the EVN.

Some of the new antennas are extending the high frequency capability of the EVN in the 22 and 43 GHz range. And not only sensitivity but certainly also image fidelity is improving with a larger number of antennas. In a number of other countries there are ambitions to construct telescopes that can be of great value to the *uv*-coverage of the existing facilities. In various SKA programs, relatively cheap, high accuracy small dishes are developed; it is an interesting option to consider small groups of these antennas (e.g. 6 x 12m), as future VLBI stations.

With such a system in place, the VLBI of the future could also be much more flexible and ready to observe at any time with the most optimal array, never compromising the optimal scientific return, yet providing immediate feedback as well. This is the VLBI facility that future users, SKA users, will want to use.